# Gradient Residual Stress in Transferred Thin-Film Lithium Niobate and Its Compensation Using Periodically Poled Piezoelectric Bilayers

Byeongjin Kim, *Student Member, IEEE*, Ian Anderson, *Student Member, IEEE*, Tzu-Hsuan Hsu, *Member, IEEE*, and Ruochen Lu, *Senior Member, IEEE*

*Abstract*—In this work, we experimentally investigate the gradient stress ($\sigma_1$) in 128° Y-cut transferred thin film lithium niobate (TFLN) films with thicknesses from 100 to 460 nm using cantilever curvature analysis. The results reveal a strong dependence of $\sigma_1$ on both crystallographic orientation and film thickness, with stress-free orientations at ~55° and ~125° for 220–460 nm films, shifting to ~20° and ~160° for 100 nm films. The extracted normalized $\sigma_1$ ranges from −0.1 to 3.4 MPa/nm (100 nm), −0.8 to 0.34 MPa/nm (220 nm), and −0.12 to 0.08 MPa/nm (460 nm), indicating a pronounced thickness-dependent through-thickness stress gradient. Finite element simulations show excellent agreement with the measurements, validating the curvature-based extraction method and confirming that $\sigma_1$ originates from an orientation-dependent residual stress gradient. To mitigate this effect, a bilayer TFLN structure with opposite crystallographic orientations, forming a periodically poled piezoelectric film (P3F), is investigated, enabling partial cancellation of $\sigma_1$. A 90/110 nm P3F bilayer reduces the equivalent normalized $\sigma_1$ to −0.4 to −0.04 MPa/nm, resulting in significantly reduced deformation. These results establish gradient stress engineering through orientation, thickness, and bilayer design as an effective strategy for achieving mechanically stable and scalable TFLN microelectromechanical systems (MEMS) devices.

## I. INTRODUCTION

RESIDUAL stress plays a critical role in the mechanical reliability and performance of thin-film micro- and nano-electromechanical systems (MEMS and NEMS), radio frequency (RF) devices, and optical components. Excessive residual stress can lead to structural deformation and failure of suspended structures during release, resulting in break, collapse, or buckling, as illustrated in Fig. 1. More generally, residual stress in thin-film systems can give rise to various mechanical instabilities and structural failures. Tensile residual stress often results in film cracking [1], [2], while compressive stress can induce delamination, buckling, or blistering of multilayer stacks [3], [4], [5]. Beyond mechanical failure, residual stress distributions can influence adhesion, fracture toughness [6], and the resonant behavior of microstructures [7]. At the same time, controlled stress has been intentionally utilized as a design parameter to enhance device functionality. Engineered stress has been shown to improve electrical conductivity [8], modify dielectric permittivity [9], enhance piezoelectric response [10], tailor magnetic anisotropy [11], and improve acoustic resonator performance [12]. These examples highlight that residual stress is not merely an unwanted byproduct of fabrication but a physical parameter that can be harnessed to optimize device performance.

Lithium niobate (LN) is a widely used piezoelectric and electro-optic material with strong elastic anisotropy and electromechanical coupling. In recent years, transferred thin-film lithium niobate (TFLN) has emerged as an enabling platform for next-generation microsystems due to its compatibility with nanofabrication and heterogeneous integration. Suspended structures fabricated in TFLN have enabled a wide range of devices including RF resonators [13], filters [14], photonic circuits [15], [16], and piezoelectric MEMS based sensors [17]. In addition, cantilever-type piezoelectric structures have been explored for sensing applications [18], piezoelectric energy harvesting [19], and optical beam steering [20],where the mechanical stability and curvature of suspended beams directly influence device performance.

Residual stress in TFLN has been investigated in several contexts. Previous studies have primarily focused on the mean residual stress ($\sigma_0$) in LN thin films rather than its spatial distribution. In epitaxial LN films, $\sigma_0$ arises from substrate clamping during cooling, which constrains the in-plane thermal expansion of the film and produces biaxial in-plane stress [21]. Similar stress has been reported in bonded LN/Si hybrid wafers due to the thermal expansion mismatch between LN and silicon during bonding and annealing [22]. Various approaches have been explored to mitigate this mean stress, including wafer thickness optimization [21], wafer-bow engineering [23], and low-temperature bonding processes [22]. Despite these efforts, residual stress is typically treated as a uniform in-plane stress averaged across the film thickness.

This work was supported by the DARPA Chemistries and monoLayers for Anti-aging Kinematics (CLOAK) program and by a fellowship from the Ministry of National Defense of the Republic of Korea. Any opinions, findings, conclusions, or recommendations expressed in this material are those of the author(s) and do not necessarily reflect the views of the Defense Advanced Research Projects Agency (DARPA).

The authors are with the Electrical and Computer Engineering Department, University of Texas at Austin, Austin, TX 78712 USA (e-mail: c18263@utexas.edu)

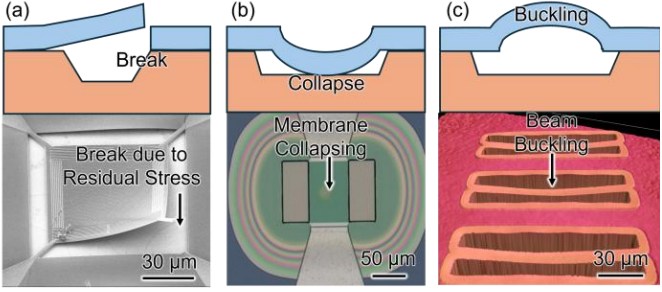

Fig. 1. Stress-induced failure modes in released thin-film piezoelectric MEMS: (a) membrane fracture from stress gradient, (b) membrane collapse, and (c) film buckling.

Recent work has indicated orientation-dependent deformation in released TFLN cantilever structures across different film thicknesses, suggesting the presence of anisotropic and gradient residual stress ($\sigma_1$) in the film [24]. However, practical strategies to actively manage or cancel such $\sigma_1$ in TFLN MEMS structures have not yet been demonstrated, and experimental observations have not been systematically validated through physics-based simulations.

This paper is extended from [24]. In this work, we experimentally investigate $\sigma_1$ in 128° Y-cut TFLN films with thicknesses ranging from 100 to 460 nm by analyzing the curvature of released cantilever structures. The measurements reveal a pronounced orientation-dependent stress behavior with stress-free orientations at ~55° and ~125° for 220–460 nm films, shifting to ~20° and ~160° for 100 nm film. The extracted normalized $\sigma_1$ ranges from −0.1 to 3.4 MPa/nm for 100 nm film, −0.8 to 0.34 MPa/nm for 220 nm films, and −0.12 to 0.08 MPa/nm for 460 nm films, indicating a strong thickness dependence of the gradient stress.

Building on these observations, we further demonstrate an approach to actively manage and cancel $\sigma_1$ using a bilayer thin-film structure consisting of two LN layers with opposite crystallographic orientations, forming a periodically poled piezoelectric film (P3F) bilayer. Such bilayer configurations, which have recently emerged in RF acoustic resonator and filter technologies [25], [26], enable effective stress compensation between the two layers. In our experiments, a 90 nm / 110 nm P3F 128° Y-cut bilayer reduces the equivalent normalized $\sigma_1$ from −0.8 to 0.34 MPa/nm in a 220 nm single-crystal film to approximately −0.4 to −0.04 MPa/nm, resulting in significantly reduced cantilever curvature. Finite-element simulations performed in COMSOL show excellent agreement with the experimental measurements, confirming the stress-compensation mechanism. These results establish gradient stress engineering and bilayer design as practical strategies for achieving mechanically stable and scalable TFLN MEMS devices.

## II. MODEL AND EXTRACTION METHOD

A uniaxial residual stress field in a thin film can be approximated by a first-order expansion through the film thickness [27]:

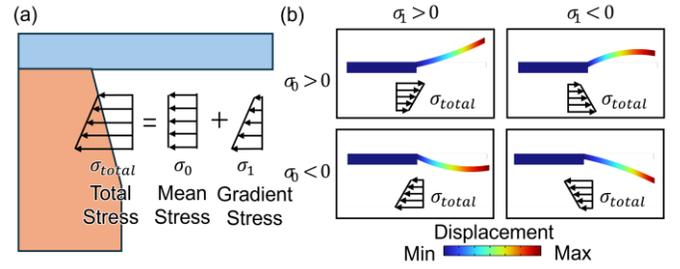

Fig. 2. Residual stress components in a TFLN cantilever: (a) decomposition into mean stress ($\sigma_0$) and gradient stress ($\sigma_1$); (b) FEA-predicted post-release beam shapes produced by these stresses (color indicates displacement).

$$\sigma_{total} = \sigma_0 + \sigma_1 \left( \frac{y}{h/2} \right) \quad (1)$$

where $h$ is the film thickness and $y \in (-h/2, h/2)$ is the thickness coordinate, defined such that $y = +h/2$ corresponds to the top surface and $y = -h/2$ corresponds to the bottom surface. With this truncation, the stress can be separated into a uniform (mean) component $\sigma_0$ and a linear gradient component $\sigma_1$. The $\sigma_0$ component is symmetric about the mid-plane, while $\sigma_1$ represents a stress gradient in the thickness direction. In typical thin-film-on-substrate systems, $\sigma_0$ is commonly associated with the coefficient of thermal expansion (CTE) mismatch between the film and the substrate [28], [29], whereas $\sigma_1$ reflects depth-dependent effects such as ion-implantation damage, bonding-induced defects, or differential relaxation during chemical-mechanical polishing (CMP) [30], as well as interface-driven stress gradients in multilayer or heterogeneous structures [31]. In addition to these factors, $\sigma_1$ can also arise from non-uniform thermal strain across the film thickness, driven by anisotropic thermal expansion in LN [32] during cooldown and the resulting differential stress relaxation between the top and bottom of the film.

To experimentally distinguish these two contributions, we adopt the cantilever-based methodology in [33], which decomposes the out-of-plane deflection profile of released beams into $\sigma_0$ and $\sigma_1$ components, as shown in Fig. 2(a). Finite element analysis (FEA), shown in Fig. 2(b), further illustrates how positive and negative $\sigma_0$ and $\sigma_1$ produce distinct deformation signatures. In particular, a positive $\sigma_1$ results in a concave-up beam profile, while a negative $\sigma_1$ leads to a concave-down curvature. In contrast, $\sigma_0$ primarily introduces a rigid-body rotation of the cantilever, where a positive $\sigma_0$ causes upward rotation near the fixed end, and vice versa. Based on this framework, the corresponding stress components can be extracted quantitatively for cantilevers following the stress–curvature calibration described in [34].

$$\sigma_1 = \frac{Eh}{2R} \quad (2)$$

where $E$ is the Young's modulus and $R$ is the signed radius of curvature. A positive $R$ corresponds to a concave-up beam profile, while a negative $R$ corresponds to a concave-down curvature. Because LN is anisotropic, $E$ must be treated as a directional modulus. After rotating the stiffness tensor into the 128° Y-cut frame [35], the effective modulus can be obtained from the rotated compliance element $S'_{11}$, which corresponds to strain under uniaxial loading along the beam direction. Thus,

$$E(\theta) = 1/S'_{11}(\theta) \qquad (3)$$

where $\theta$ is the in-plane orientation angle of the cantilever. Using this substitution, (2) remains valid with $E$ replaced by $1/S'_{11}$ [36], enabling direct evaluation of stress components for cantilevers oriented along any in-plane direction in 128° Y-cut TFLN.

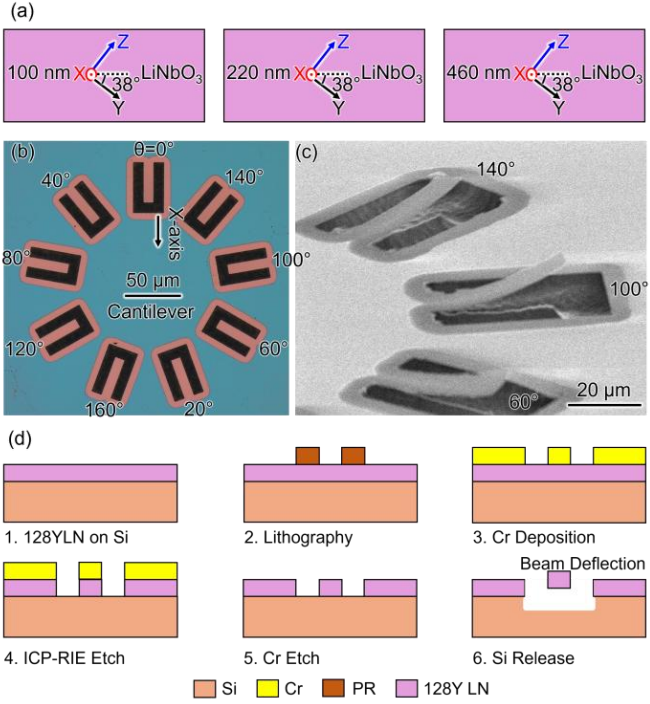

Fig. 3. TFLN cantilever array and film structure: (a) schematic of single-layer films (100, 220, and 460 nm) with crystallographic orientations; (b) optical image of cantilever array with θ = 0°–160° in 20° steps; (c) SEM image of released cantilevers; (d) fabrication process flow.

We note that the magnitude of $\sigma_1$ can vary with in-plane orientation in 128° Y-cut TFLN. This variation does not arise solely from elastic anisotropy, which primarily governs orientation-dependent stiffness, but instead reflects an underlying through-thickness residual-stress gradient. Such gradients can originate from multiple depth-dependent mechanisms, including processing-induced damage, interfacial effects, and non-uniform relaxation across the film thickness. Based on our measurements, however, we attribute the dominant contribution to anisotropic thermal expansion in LN, which induces direction-dependent thermal strain during cooldown and leads to asymmetric stress relaxation between the top and bottom portions of the film. Consequently, $\sigma_1$ reflects the through-thickness residual stress profile rather than purely elastic effects.

In isotropic thin films, both $\sigma_0$ and $\sigma_1$ can be reliably extracted from released beam profiles using curvature–rotation relationships [27]. However, this decomposition does not directly carry over to anisotropic TFLN. Unlike isotropic materials such as $SiO_2$, the total beam rotation in LN is strongly influenced by anisotropic bending stiffness and stress localization near the release junction, as shown in Fig. 8(a). As a result, the rotation can vary significantly with beam width and length, making $\sigma_0$ highly geometry-dependent and difficult to extract reproducibly using curvature-based methods in 128° Y-cut TFLN. In contrast, $\sigma_1$ is determined primarily by the normalized curvature $h/R$ and is largely insensitive to beam width and length, making it a more robust metric across different film thicknesses and device geometries. Finite element simulations further validate this approach by showing strong agreement between measured and predicted curvature behavior. Therefore, in this work, we focus on $\sigma_1$, the gradient stress component, as the primary measure of in-plane residual stress anisotropy in suspended 128° Y-cut TFLN. While direct measurement of through-thickness stress profiles may be pursued in future studies using techniques such as X-ray diffraction (XRD) [37], [38] or focused-ion-beam (FIB) strain-relief methods [39], [40], such approaches are beyond the scope of this work. Here, we instead adopt a device-based methodology that provides a complementary and practical means of capturing stress anisotropy without introducing additional experimental complexity.

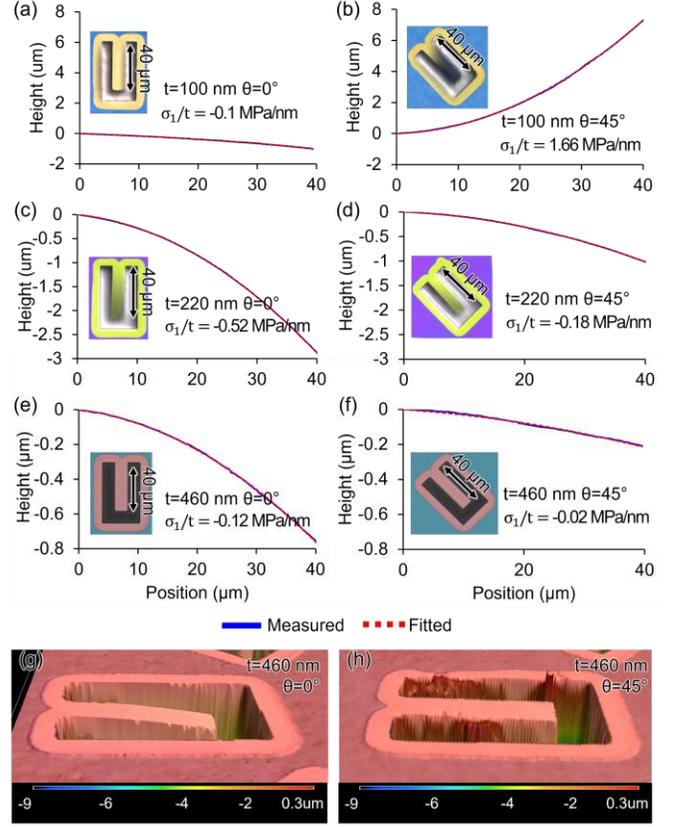

Fig. 4. Optical profilometry of released TFLN cantilevers: (a–b) 100 nm, (c-d) 220 nm, and (e-f) 460 nm films showing measured and fitted out-of-plane beam profiles with normalized gradient stress (σ₁/t) at θ = 0° and 45°; (g–h) corresponding 3D height maps for the 460 nm film (color indicates surface height).

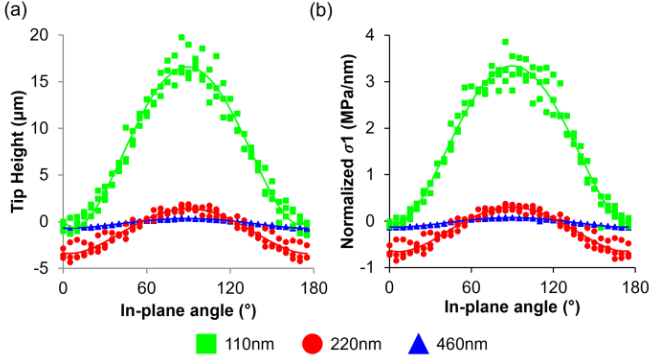

Fig. 5. Beam tip height and normalized $\sigma_1$ versus in-plane angle in TFLN cantilevers: (a–b) single-layer 128° Y-cut films with thicknesses of 110, 220, and 460 nm.

## III. FABRICATION

Micromachined cantilever arrays were fabricated from 128° Y-cut TFLN on Si substrates provided by NGK Corporation. In addition to single-layer films with thicknesses of 100 nm, 220 nm, and 460 nm, a bilayer stack was investigated to examine the effect of multilayer structures on residual stress behavior. The bilayer is based on a P3F configuration with opposite crystallographic orientations in the two layers, which ideally enables symmetric stress compensation. In this work, the bilayer consists of 90 nm (top) and 110 nm (bottom) LN layers, reflecting slight thickness variation in the as-fabricated stack. The crystallographic configurations of both single-layer and bilayer structures are illustrated in Fig. 3(a).

Each cantilever set consisted of beams with in-plane orientations ranging from θ = 0° to 160° in 20° increments, as shown in Fig. 3(b). To achieve finer angular resolution, four identical sets were patterned with a 5° rotation offset between them, resulting in a combined array spanning θ = 0° to 175° in 5° increments (36 orientation groups per film thickness). This layout enables systematic mapping of orientation-dependent stress behavior across the wafer. Representative released cantilevers at different orientations are shown in the SEM image in Fig. 3(c), illustrating the strong orientation dependence of the beam curvature.

The fabrication process is summarized in Fig. 3(d). First, the cantilever patterns were defined using maskless photolithography. Following lithography, a chromium (Cr) layer was deposited by electron-beam evaporation to serve as a hard mask for subsequent dry etching. A lift-off process was then performed to define the Cr etch mask, leaving Cr only in the desired cantilever regions. Using this Cr hard mask, the exposed 128° Y-cut LN layer was etched by inductively coupled plasma reactive ion etching (ICP-RIE). A low-temperature ICP-RIE process was employed to minimize additional thermal residual stress during etching. After LN patterning, the Cr hard mask was removed using a selective Cr wet etchant, ensuring that no residual masking material remained on the device surface. Finally, the suspended cantilever structures were released by isotropic silicon etching using XeF₂. This dry release process selectively removes the underlying Si substrate beneath the patterned LN without damaging the thin film, resulting in fully suspended cantilever beams.

## IV. MEASUREMENT OF RESIDUAL STRESS IN THIN-FILM LN

The resulting cantilever curvature was measured using a laser confocal optical profilometer, presented in Fig. 4 (a)-(f). The measured beam deflection profiles were fitted with a quadratic model following the method in [27], enabling the extraction of curvature and rotation parameters. The fitting procedure incorporates the anisotropic elastic response of LN through the directional modulus defined in (3). In addition, three-dimensional height maps of the beams were obtained to validate the curvature fitting results, as shown in Fig. 4 (g) (h). For each orientation group, measurements were performed at three different chip locations to verify reproducibility and reduce the influence of local process variation. Together, this experimental platform enables systematic extraction of orientation-dependent gradient stress across multiple film thicknesses and structural configurations, providing a comprehensive mapping of in-plane residual stress anisotropy in TFLN.

Fig. 5 summarizes the extracted beam tip heights together with the corresponding normalized $\sigma_1$ across different film thicknesses and orientations. Here, $\sigma_1$ is defined as the linear gradient component of the residual stress, corresponding to the stress difference between the film mid-plane and either the top or bottom surface [27]. Because this definition inherently depends on the film thickness, direct comparison of $\sigma_1$ across films of different thicknesses can be misleading. Therefore, we normalize $\sigma_1$ by the film thickness to obtain a thickness-independent metric (in units of MPa/nm), enabling a consistent comparison of gradient stress across different TFLN films.

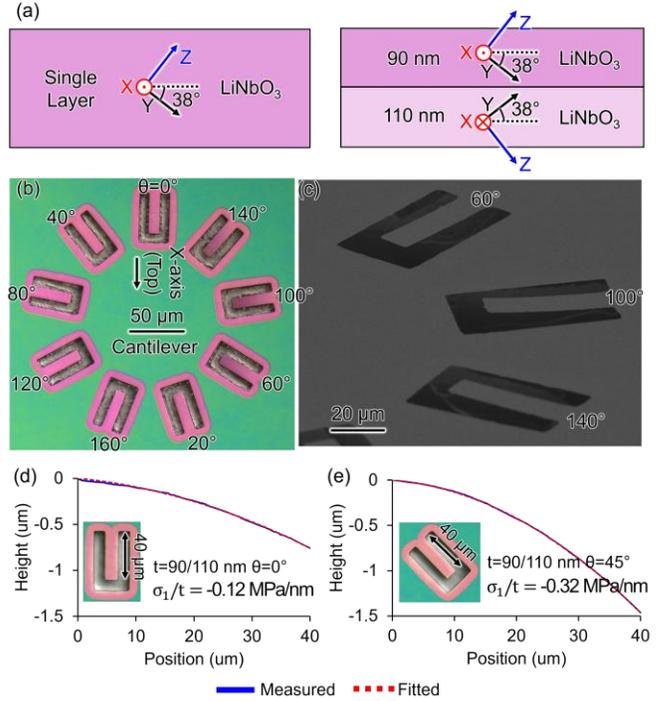

Fig. 6. Bilayer TFLN cantilever array and structure: (a) schematic of single-layer and 90/110 nm bilayer configurations with crystallographic orientations; (b) optical image of the cantilever array (θ = 0°–160°); (c) SEM image of released cantilevers; (d–e) measured and fitted out-of-plane profiles of bilayer cantilevers at θ = 0° and 45°, showing normalized gradient stress (σ₁/t).

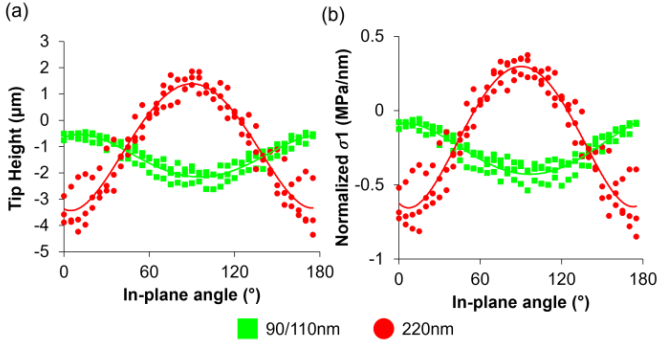

Fig. 7. Beam tip height and normalized $\sigma_1$ versus in-plane angle in TFLN cantilevers: (a–b) comparison between a 220 nm single-layer and a 90/110 nm bilayer 128° Y-cut film.

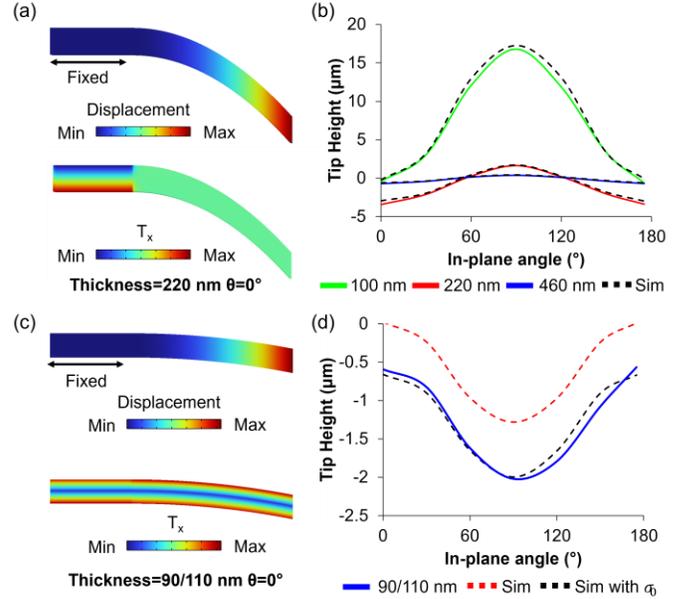

Fig. 8. Simulated displacement and $T_x$ for (a) single-layer (220 nm) and (c) bilayer (90/110 nm) cantilevers at $\theta = 0°$; measured and simulated beam tip height profiles for (b) single-layer and (d) bilayer cantilevers.

For both 460 nm and 220 nm single-layer films, $\sigma_1$ exhibits a strong orientation dependence, crossing zero near ~55° and ~125°. These stress-free orientations correspond closely to the angles at which the cantilevers appear nearly flat, with a deviation within ~5°, indicating that the residual mean stress $\sigma_0$ plays a relatively minor role in determining the beam shape and remains comparatively small. Instead, $\sigma_1$ serves as the dominant descriptor of in-plane residual stress in these suspended cantilevers. Quantitatively, the extracted $\sigma_1$ ranges from −0.12 to 0.08 MPa/nm for the 460 nm film and from −0.8 to 0.34 MPa/nm for the 220 nm film, confirming a clear thickness dependence of the gradient stress. As the film becomes thinner, the magnitude of $\sigma_1$ increases significantly, indicating a stronger through-thickness stress asymmetry.

This trend becomes more pronounced in the ultra-thin 100 nm film, where the stress-free orientations shift to ~20° and ~160°. The extracted $\sigma_1$ spans a much wider range from −0.1 to 3.4 MPa/nm, and the cantilevers exhibit a noticeably larger upward curvature compared to thicker films, corresponding to a stronger positive $\sigma_1$ (i.e., a more tensile upper surface). These observations highlight the increasing dominance of gradient stress in ultra-thin TFLN and its strong dependence on both thickness and crystallographic orientation.

## V. RESIDUAL STRESS COMPENSATION IN P3F LN BILAYERS

P3F LN has recently emerged as a multilayer approach to enhancing device performance in RF resonators and filters by incorporating multiple crystallographic orientations into a single structure [26], [41], [42]. In these configurations, alternating layer orientations, typically achieved through twofold (180°) rotations, modify the effective piezoelectric coupling and enable constructive interaction with acoustic modes that would otherwise be partially canceled in single-layer systems [43], [44]. Beyond RF applications, such multilayer architectures also provide additional degrees of freedom for tailoring mechanical and electromechanical properties.

In this work, we leverage this concept to engineer residual stress in 128° Y-cut TFLN by employing a P3F bilayer structure, as illustrated in Fig. 6 (a). The bilayer stack (90/110 nm) is designed to introduce opposing gradient stress components between the top and bottom layers. Cantilever arrays with varying in-plane orientations are fabricated and characterized using the same methodology established for single-layer films, including optical imaging, SEM observation, and profilometry-based curvature fitting [Fig. 6 (b)–(e)]. These measurements enable direct extraction of the normalized $\sigma_1$ in the bilayer system, allowing for a consistent comparison with the single-layer results presented in Section IV.

Fig. 7 compares the stress behavior between a 220 nm single-layer film and a bilayer 90/110 nm TFLN structure. The bilayer device exhibits a substantially reduced $\sigma_1$ across all in-plane orientations, with the extracted range reduced to approximately −0.4 to −0.04 MPa/nm, compared to −0.8 to 0.34 MPa/nm for the single-layer counterpart. Notably, although the bilayer has a total thickness of 200 nm, a single-layer film of this thickness would be expected to exhibit a larger $\sigma_1$ based on the observed thickness-dependent trend, in which thinner films show increased gradient stress. The reduced $\sigma_1$ in the bilayer, therefore, indicates effective stress cancellation in the P3F structure.

This cancellation arises from the opposing crystallographic orientations of the two LN layers, which generate gradient stresses of opposite sign that partially compensate. As a result, the overall curvature of the cantilevers is significantly reduced. However, the bilayer devices are not completely flat, and the $\sigma_1$ curve exhibits a concave-up trend, in contrast to the convex behavior observed in single-layer films. This residual asymmetry is attributed to the thickness imbalance between the two layers, where the thicker bottom layer contributes a dominant compressive component.

Overall, these results confirm that the normalized $\sigma_1$ is strongly dependent on both thickness and crystallographic orientation in single-layer TFLN, whereas bilayer structures provide an effective means of suppressing this stress. These findings collectively demonstrate that careful control of orientation, thickness, and layer configuration can mitigate stress-induced deformation, enabling more reliable and scalable fabrication of suspended MEMS devices in TFLN.

## VI. Validation of Stress Cancellation in P3F LN

To validate the extracted $\sigma_1$ values, FEA was performed using the stress model defined in (1)–(3). The measured $\sigma_1$ was directly applied as an initial stress condition in the TFLN layer, thereby enabling it to serve as a residual stress field after substrate release. In the simulations, one end of the cantilever was fixed to the bottom surface, while the remaining structure was left free to mimic a released boundary condition. The initial stress was assigned according to (1), where $\sigma_1$ denotes the through-thickness stress gradient responsible for beam curvature, and $\sigma_0$ denotes a uniform mean stress. This approach enables verification that the extracted $\sigma_1$ accurately reproduces the observed beam deformation and provides insight into the internal stress distribution via the in-plane stress component ($T_x$).

Fig. 8 (a) shows the simulated displacement and $T_x$ for a 220 nm single-layer cantilever at $\theta = 0°$. The simulation confirms the strong stress localization near the release junction, consistent with the interpretation discussed in the design section. This behavior explains why beam rotation is strongly influenced by local stress concentration and anisotropic stiffness, rendering $\sigma_0$ highly geometry-dependent and difficult to extract reliably using curvature-based methods. Simulations were performed from 0° to 175° in 5° increments for 100-, 220-, and 460-nm films, and the resulting beam tip heights are shown in Fig. 8 (b). The simulated results (dotted lines) show excellent agreement with the measured data (solid lines), confirming that the extracted $\sigma_1$ accurately captures the deformation behavior. The small deviations are attributed to the omission of $\sigma_0$ in the simulation.

In the bilayer case, $\sigma_1$ was not measured directly for the individual layer thicknesses. Instead, it was estimated from the 100 nm and 220 nm data using a thickness-dependent interpolation. This approach reflects the experimentally observed increase in $\sigma_1$ for thinner films (Fig. 5) by assigning them a higher weight. The interpolated $\sigma_1$ values were then used as input for each layer in the bilayer simulation. Fig. 8 (c) shows the simulated displacement and $T_x$ for a bilayer cantilever (90/110 nm) at $\theta = 0°$. Unlike the single-layer case, the bilayer structure exhibits opposite stress distributions in the top and bottom layers due to their reversed crystallographic orientations, resulting in cancellation of the gradient stress. Consequently, the net curvature is significantly reduced. However, complete cancellation is not achieved because the bottom layer is thicker than the top layer, leading to a residual compressive contribution that biases the deformation. The comparison between measured and simulated beam tip height profiles for the bilayer is shown in Fig. 8 (d). The simulation without $\sigma_0$ (red dotted line) captures the overall trend but shows a noticeable offset from the measured data (solid line). Introducing an empirical mean stress offset of $\sigma_0 = -2$ GPa (black dotted line) significantly improved the agreement. We note that this value is larger than the reported mean stress of ~155 MPa in thick (~5 μm) LNOI films [22], but remains consistent with prior reports of GPa-level stress in submicron LN layers (e.g., ~1.05 GPa for 270 nm LN on sapphire) [21], indicating a strong thickness dependence of $\sigma_0$. In addition, the thermal expansion mismatch between LN and sapphire ($\alpha \approx 14.4$–$15.9 \times 10^{-6}$/K for LN [45] vs. $8.11 \times 10^{-6}$/K for sapphire [46]) is larger than that between LN and Si ($\alpha \approx 3.57 \times 10^{-6}$/K [46]), further supporting the possibility of elevated mean stress in thin-film LN systems. While the magnitude of $\sigma_0$ is therefore physically reasonable, its exact value is obtained empirically to match the measured deformation and should be interpreted as an effective fitting parameter rather than a directly measured material property. The remaining mismatch in the 100°–175° range is attributed to local variations in layer thickness across the chip.

Overall, the strong agreement between measurement and simulation confirms the validity of the cantilever curvature method for extracting $\sigma_1$. These results suggest that the observed $\sigma_1$ originates from an orientation-dependent through-thickness residual gradient stress. A likely source is anisotropic thermal expansion of LN during cooldown, coupled with asymmetric stress relaxation between the film's top and bottom portions. In addition, the observed stress reduction in the bilayer structure confirms that opposing crystallographic orientations can effectively mitigate gradient stress.

On the practical application of P3F LN in acoustic devices, it is important to note that the bilayer configuration in 128° Y-cut LN, despite consisting of oppositely oriented crystal layers, can preserve the effective piezoelectric response by maintaining constructive electromechanical interaction across the multilayer. In particular, the alternating crystallographic orientations reverse the sign of the piezoelectric coefficients, preventing cancellation of the induced electric field [41]. As a result, the P3F bilayer can achieve the same electromechanical coupling ($k^2$) while also providing additional benefits, such as spurious-mode suppression, compared to single-layer platforms [43], [44]. At the same time, the bilayer configuration provides an additional mechanical advantage not available in conventional single-layer designs. As demonstrated in this work, the opposing orientations generate gradient stress components of opposite sign, leading to partial cancellation of $\sigma_1$ and a significant reduction in structural deformation. This stress compensation is achieved without degrading piezoelectric performance, providing a pathway to simultaneously address both mechanical stability and acoustic performance in TFLN devices.

## VII. Conclusion

In this work, we investigated residual gradient stress in 128° Y-cut TFLN using cantilever curvature. The results show a strong dependence of $\sigma_1$ on crystallographic orientation and film thickness, with stress-free orientations and significantly larger gradient stress in thinner films. The extracted normalized $\sigma_1$ ranges from $-0.1$ to $3.4$ MPa/nm (100 nm), yielding −0.8 to 0.34 MPa/nm (220 nm) and −0.12 to 0.08 MPa/nm (460 nm), demonstrating a pronounced thickness dependence. A bilayer structure with opposite crystallographic orientations effectively suppresses $\sigma_1$ via stress cancellation, reducing it to −0.4 to −0.04 MPa/nm in a 90/110-nm bilayer. Finite element simulations show excellent agreement with the measurements, validating the curvature-based extraction of $\sigma_1$ and supporting that it originates from an orientation-dependent through-thickness residual stress gradient, likely driven by anisotropic thermal expansion and asymmetric stress relaxation. These results establish gradient-stress engineering as a practical strategy for achieving mechanically stable, scalable TFLN MEMS devices.

ACKNOWLEDGMENT

The authors would like to thank Dr. Weileun Fang, Dr. Chao-Lin Cheng, and Dr. Sunil Bhave for helpful discussions.

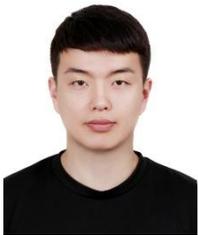

**BYEONGJIN KIM** (Student Member, IEEE) received the B.S. degree in applied physics from Korea Military Academy, Seoul, South Korea, in 2019. He is currently pursuing the M.S. degree in Electrical and Computer Engineeringath the University of Texas at Austin. Since 2019, he has been serving as an Army officer in the Republic of Korea and is supported by a scholarship from the Ministry of National Defense. His current research focuses on scandium aluminum nitride (ScAlN) bulk acoustic wave (BAW) resonators and the impact of residual stress on device performance. His research interests include piezoelectric MEMS, acoustic resonators, and stress engineering in thin-film devices.

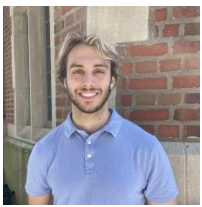

**IAN ANDERSON** (Student member, IEEE) received the B.S. degree in Electrical and Computer Engineering from the Ohio State University, Columbus, OH, USA, in 2023. He is currently a PhD student in Electrical and Computer Engineering at the University of Texas at Austin, Austin, TX, USA. His current work focuses on design and fabrication of thin film bulk acoustic resonators for infrared sensors, and work on high frequency phononic combs. He also has further research interests and projects in solidly mounted acoustic devices and acousto-optic modulators. He is a recipient of the NASA Space Technology Graduate Research Opportunity (NSTGRO) Fellowship.

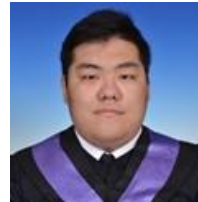

**TZU-HSUAN HSU** (Member, IEEE) is a Postdoctoral Researcher with the Department of Electrical and Computer Engineering at The University of Texas at Austin. He received the B.S. degree in power mechanical engineering from National Tsing Hua University (NTHU), Taiwan in 2017 and the M.S. degree from the Institute of NanoEnigneering and MicroSystems, NTHU, Taiwan in 2019, respectively. He later received the Ph.D. degree in power mechanical engineering from NTHU, Hsinchu, Taiwan, in 2024. His research focuses on the development of advanced piezoelectric RF acoustic devices targeting the next generation wireless signal processing. He received the Scholarship Pilot Program to Cultivate Outstanding Doctoral Students from Ministry of Science and Technology of Taiwan in 2020, the CTCI Foundation Science and Technology Research Scholarship in 2021. He received the Best Student Paper Award at the 2020 Joint Conference of the IEEE International Frequency Control Symposium, and the IEEE International Symposium on Applications of Ferroelectrics (IFCS-ISAF 2020).

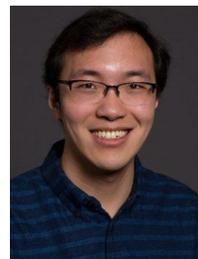

**RUOCHEN LU** (Senior Member, IEEE) is an Assistant Professor in the Department of Electrical and Computer Engineering at The University of Texas at Austin. He received the B.E. degree with honors in microelectronics from Tsinghua University in 2014, and the M.S. and Ph.D. degrees in electrical and computer engineering from the University of Illinois at Urbana-Champaign in 2017 and 2019, respectively. His research focuses on MEMS and thin-film piezoelectric microsystems for signal processing, sensing, and computing. His group develops chip-scale acoustic and electromagnetic devices spanning RF to the mm-wave/sub-THz regime, including resonators, filters, ultrasound transducers, and hybrid microsystems that interface acoustics with electronics, photonics, magnetics, and power conversion. He received the IEEE MTT-S Microwave Award in 2022, the NSF CAREER Award and IEEE Ultrasonics Early Career Investigator Award in 2024, and the Junior Faculty Excellence in Teaching Award from UT Austin in 2024. He is an Associate Editor of the IEEE Journal of Microelectromechanical Systems, IEEE Transactions on Ultrasonics, IEEE Journal of Microwaves, and IEEE Electron Device Letters.